\documentclass{elsart}
\usepackage{graphicx}
\usepackage{amssymb}

\begin{document}

\begin{frontmatter}

\title{Correlation between diffusion and coherence in
Brownian motion on a tilted periodic potential}

\author{E. Heinsalu},
\ead{ehe@ut.ee}
\author{R. Tammelo},
\author{T. \"{O}rd\corauthref{cor}}
\corauth[cor]{corresponding author}
\ead{teetord@ut.ee}

\address{Institute of Theoretical Physics, University
of Tartu,\\ 4 T\"{a}he Str., 51010 Tartu, Estonia}

\begin{abstract}
The paper studies the overdamped motion of Brownian particles in a
tilted sawtooth potential. The dependencies of the diffusion
coefficient and coherence level of Brownian transport on
temperature, tilting force, and the shape of the potential are
analyzed. It is demonstrated that at low temperatures the
coherence level of Brownian transport stabilizes in the extensive
domain of the tilting force where the value of the P\'{e}clet
factor is $Pe=2$. This domain coincides with the one where the
enhancement of the diffusion coefficient \textit{versus} the
tilting force is the most rapid. The necessary and sufficient
conditions for the non-monotonic behaviour of the diffusion
coefficient as a function of temperature are found. The effect of
the acceleration of diffusion by bias and temperature is
demonstrated to be very sensitive to the value of the asymmetry
parameter of the potential.
\end{abstract}

\end{frontmatter}

In recent years the anomalous properties of thermal diffusion in
tilted periodic potentials have been considered in a number of
papers \cite{constantini,reimann2001,reimann2002,lindner2001,dan}.
The effect of the giant amplification of diffusion by bias with
respect to free diffusion was observed
\cite{reimann2001,reimann2002} and the non-monotonic behaviour of
the diffusion coefficient as a function of temperature was found
\cite{lindner2001}. The influence of spatially modulated friction
on diffusion and coherent transport in a tilted washboard
potential was investigated in Ref. \cite{dan}, and similar effects
caused by spatially periodic temperature were discussed in Ref.
\cite{lindner2002}.

In the present contribution we study the diffusion and coherence
of overdamped Brownian particles in a biased sawtooth potential in
the space of the tilting force $F$, temperature $T$, and asymmetry
parameter $k$ of the potential. We ascertain the characteristic
features of Brownian transport in the system under consideration
and obtain the relevant conditions for non-monotonic dependence of
the diffusion coefficient on temperature.

The equation of one-dimensional overdamped motion of a Brownian
particle in dimensionless units reads
\begin{equation}
\frac{dx(t)}{dt}=-\frac {dV(x)}{dx}+\xi(t)\;,
\end{equation}
\vspace{-13pt}

where $V(x)$ is a tilted periodic potential and $\xi(t)$ is
Gaussian white noise.

Applying the general approach developed in Refs.
\cite{reimann2001,reimann2002} to a piecewise linear potential of
the period $L=1$, we obtain by means of cumbersome calculations
the following algebraic expressions for the current and diffusion
coefficient ($0<k<1$, $k=0.5$ corresponds to the symmetric
potential; $F\geq0$, while above the critical tilt, $F>F_{cr}=1$,
the potential has no local minima):
\begin{equation}
\langle\dot{x}\rangle=\frac{\varphi_{0}}{Z}\;,\; \; \; \;
D=\frac{TY}{Z^{3}}\;,
\end{equation}
\vspace{-13pt}

where

\vspace{-18pt}
\begin{eqnarray}
Z & = & \left(\frac{k}{a}-\frac{1-k}{b}\right)\varphi_{0}+
T\left(\frac{1}{a}+\frac{1}{b}\right)^{2}\varphi_{a}\varphi_{b}\;, \\
Y & = &
\left(\frac{k}{a^{3}}-\frac{1-k}{b^{3}}\right)\varphi_{0}^{3}
+3T\left(\frac{1}{a^{3}}+\frac{1}{b^{3}}\right) \left(\frac{1}{a}+
\frac{1}{b}\right)\varphi_{0}^{2}\varphi_{a}\varphi_{b} \nonumber\\
 & + & \frac{1}{2}T \left(\frac{1}{a}+ \frac{1}{b}\right)^{2}\varphi_{0}
\left[\frac{1}{a^{2}}\varphi_{a}^{2}\tilde{\varphi_{b}}-
\frac{1}{b^{2}}\varphi_{b}^{2}\tilde{\varphi_{a}}\right] \nonumber\\
& + & 2\left(\frac{1}{a}+\frac{1}{b}\right)^{2}\varphi_{0}
\left[\frac{k}{a}\varphi_{a}^{2}(1-\varphi_{b})-
\frac{1-k}{b}\varphi_{b}^{2}(1+\varphi_{a})\right] \nonumber\\
& + & T\left(\frac{1}{a}+\frac{1}{b}\right)^{3}
\left[\frac{1}{a}\varphi_{a}^{3}\varphi_{b}(1-\varphi_{b})+
\frac{1}{b}\varphi_{b}^{3}\varphi_{a}(1+\varphi_{a})\right] \; ,
\end{eqnarray}
\vspace{-25pt}

with

\vspace{-13pt}
\begin{eqnarray}
a & = & \frac{1-(1-F)k}{(1-k)k}\;, \; \; \; \;
b=\frac{1-F}{1-k}\;, \\
\varphi_{0} & = & 1-\exp\left({-\frac{F}{T(1-k)}}\right)\;, \\
\varphi_{a} & = & \exp\left({\frac{1-F}{T}}\right)-1\;,\; \; \; \;
\varphi_{b}=1-\exp\left({-\frac{1-(1-F)k}{T(1-k)}}\right)\;, \\
\tilde{\varphi_{a}} & = & \exp\left({\frac{2(1-F)}{T}}\right)-1\;,
\; \;
\;\;\tilde{\varphi_{b}}=1-\exp\left({-\frac{2[1-(1-F)k]}{T(1-k)}}\right)\;.
\end{eqnarray}
\vspace{-20pt}

A relevant quantity characterizing the coherence
level of Brownian motion is the P\'{e}clet factor,
\begin{equation}\label{Pe}
Pe=\frac{\langle \dot{x}\rangle}{D}=\frac{\varphi_{0}Z^{2}}{TY}\;.
\end{equation}
\vspace{-15pt}

On the basis of the algebraic formulae obtained, we will analyze
the behaviour of the diffusion coefficient and the P\'{e}clet
factor in the space of the system parameters.

Firstly, we consider the correlation between the coherence level
of Brownian transport and the acceleration of diffusion by the
external tilting force. The comparative plot of $D$ and $Pe$
\textit{versus} $F$ is shown in Fig.~1.
\begin{figure}[t]
\begin{center}
\includegraphics[width=0.5\linewidth]
{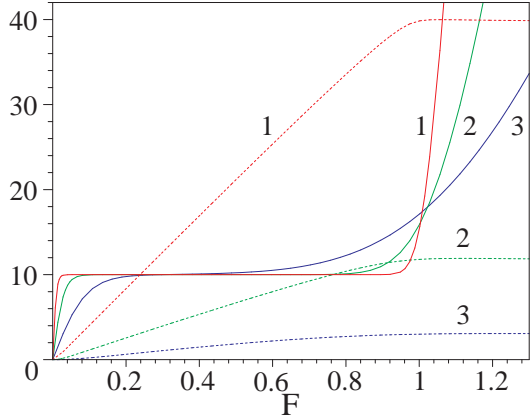} \caption{Comparison of the dependencies of the
P\'{e}clet factor and diffusion coefficient on the tilting force
for various temperatures at fixed $k=0.5$. Solid lines: $5 \times
Pe$ \textit{vs} $F$, dashed lines: $\log[D(F)/D(0)]$ \textit{vs}
$F$. Curves 1: $T=0.01$, curves 2: $T=0.03$, curves 3: $T=0.09$.}
\end{center}
\end{figure}
One can see that the function $Pe(F)$ has a point of inflection
which turns into a wide plateau at low temperatures. For the
values of $F$ from zero up to the end of the plateau, the
behaviour of $Pe(F)$ is described with great accuracy by the
expression
\begin{equation}
Pe=2\tanh \frac{F}{2T(1-k)} \, ,
\end{equation}
\vspace{-15pt}

which stems analytically from Eq. (\ref{Pe}) in the proper
approximation. We emphasize that in the same domain where $Pe=2$,
the increase of the diffusion coefficient caused by the tilt is
the most rapid, following quite exactly the law
$c^{\alpha_{1}+\alpha_{2} F}$ where $c$ and $\alpha_{1,2}$ depend
on $T$ and $k$. Consequently, in the region of parameters, where
the substantial acceleration of diffusion (and also current)
occurs, the current and the diffusion are synchronized. Note also
that the stabilization of the coherence level at the value of the
P\'{e}clet factor $Pe=2$ is a characteristic feature of the
Poisson process \cite{risken} such as the Poisson enzymes in
kinesin kinetics \cite{svoboda,schnitzer,visscher}.
\begin{figure}[t]
\begin{center}
\includegraphics[width=0.6\linewidth]
{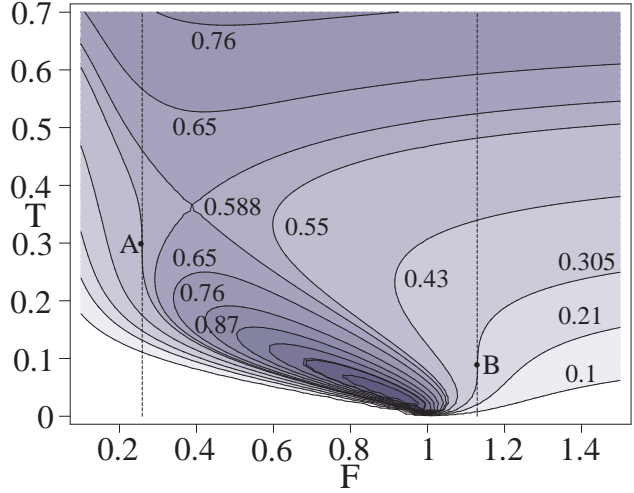} \caption{Contourplot of the surface $D=D(T,F)$
for $k=0.95$. The values of the diffusion coefficient
$D(T,F)=\textrm{const}$ are displayed. To the maximum and saddle
points of $D$ correspond, respectively, the values $F_M \approx
0.9144,$ $T_M \approx 0.0364, D_M \approx 1.3086$ and $F_S \approx
0.388, T_S \approx 0.363,$ $D_S \approx 0.588$.}
\end{center}
\end{figure}

Now let us discuss the dependence of diffusion on the system
parameters. Figure 2 presents the analytic properties of the
diffusion coefficient as a function of temperature and tilting
force, displaying the contourplot of the surface $D=D(T,F)$. The
surface exhibits two stationary points, a maximum and a saddle
point, whose coordinates are given in the figure caption. The plot
reflects fully the characteristic features of the non-monotonic
behaviour of diffusion: (i) One can observe in Fig.~2 that the
function $D(T)\!\!\! \mid_{F=\textnormal{\scriptsize{const}}}$ has
a maximum and a minimum if $F_{A}<F<F_{B}$. The maximum of $D(T)$
becomes rapidly narrower and higher as $k$ approaches unity (see
Fig.~3).
\begin{figure}[h]
\begin{center}
\includegraphics[width=0.6\linewidth]
{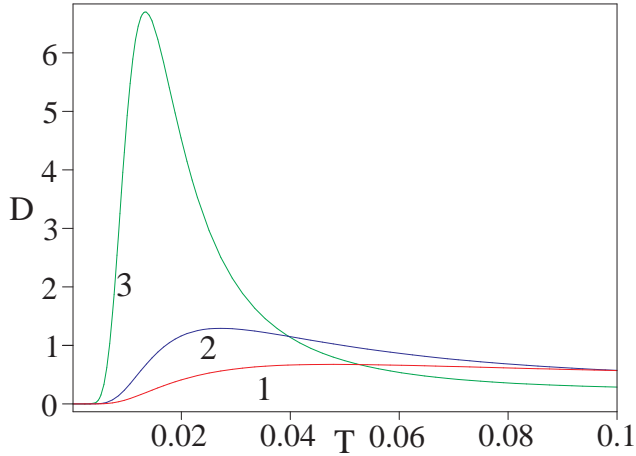} \caption{The diffusion coefficient $D$
\textit{vs} temperature $T$ at $F=0.95$ for various values of the
asymmetry parameter. Curve 1: $k=0.9$, curve 2: $k=0.95$, curve 3:
$k=0.99$.}
\end{center}
\end{figure}
\begin{figure}[t]
\begin{center}
\includegraphics[width=0.5\linewidth]
{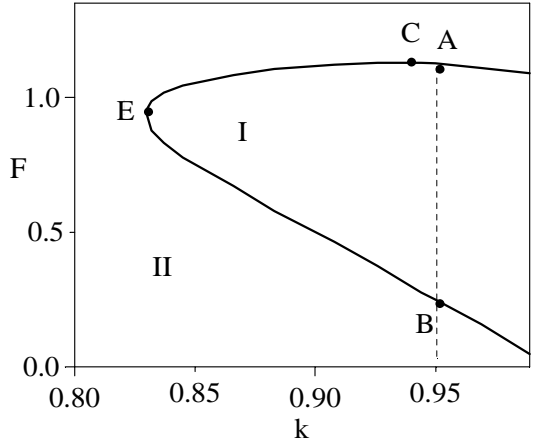} \caption{The phase-diagram in the
$(F,k)$-plane representing the regions corresponding to the
different analytical properties of the diffusion coefficient as a
function of temperature: the dependence $D(T)$ is non-monotonic in
the region I, whereas it is monotonic in the region II.}
\end{center}
\end{figure}
\begin{figure}[h]
\hspace{1.7cm}
\includegraphics[width=0.63\linewidth]
{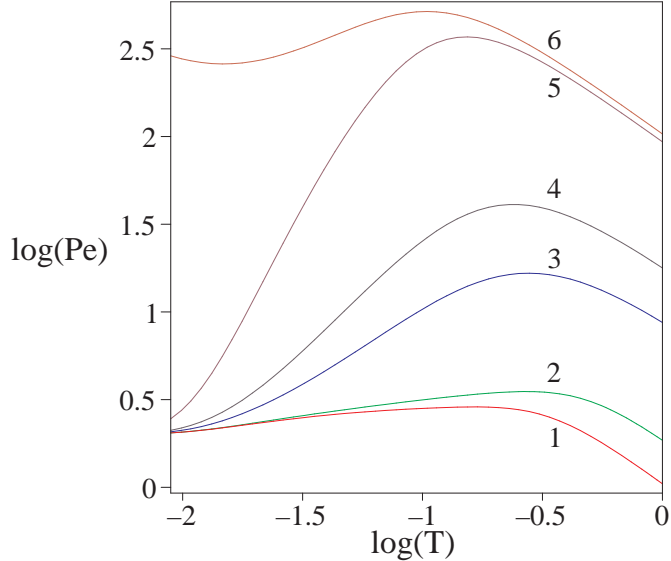} \caption{The plot of the P\'{e}clet factor
\textit{vs} temperature for various values of the asymmetry
parameter $k$ of the potential at $F=0.95$ (curves 1-5) and
$F=1.05$ (curve 6). Curve 1: $k=0.1$, curve 2: $k=0.5$, curve 3:
$k=0.9$, curve 4: $k=0.95$, curves 5, 6: $k=0.99$.}
\end{figure}
For $k<k_{E}\approx 0.8285$, the saddle point of the surface
$D(T,F)$ disappears, while
$D(T)\!\!\mid_{F=\textnormal{\scriptsize{const}}}$ is a monotonic
function of temperature, the latter property being independent of
bias. There also exists a limiting tilting force $F_{C}\approx
1.1292$. If $F>F_{C}$, the dependence
$D(T)\!\!\!\mid_{F=\textnormal{\scriptsize{const}}}$ is monotonic
for arbitrary $k$. The situation is summarized by the
phase-diagram in Fig.~4. (ii) Contrary to the dependence
$D(T)\!\!\!\mid_{F=\textnormal{\scriptsize{const}}}$, the function
$D(F)\!\!\!\mid_{T=\textnormal{\scriptsize{const}}}$ always has a
maximum, while the larger values of $k$ are more favourable for
the amplification of diffusion by bias.

The dependence of the P\'{e}clet factor on temperature for various
values of $k$ and $F$ is depicted in Fig.~5. The curves $Pe(T)$
have a maximum, which occurs also for the values of the tilt
slightly above the critical (curve 6). With a further increase of
$F$, the maximum of $Pe(T)$ disappears. As it is seen in Fig.~5,
the optimal level of Brownian transport determined by the maximal
value of the P\'{e}clet number is sensitive to the variation of
the shape of the periodic potential: the optimal level rises with
the increase of $k$. At the same time, if $k>k_{E}$, the
dependence $D(T)$ can be non-monotonic (see Fig.~4). Then the
larger values of $k$ deepen the minimum of $D(T)$, which follows
the maximum as temperature increases, enlarging by that the
P\'{e}clet number. In this sense the situation is analogous to the
results of Ref. \cite{dan} where the enhancement of the coherence
of Brownian motion in a certain temperature range due to
frictional inhomogeneity associates with the suppression of
diffusion by the same factor.

To conclude, we emphasize that there exists a significant
correlation between the acceleration of diffusion by tilting force
and the stabilization of the coherence level of Brownian motion,
resulting from the interplay of periodic potential, bias and white
noise. It seems that this phenomenon is quite universal and
manifests itself for arbitrary periodic potentials where initially
strongly suppressed transport is enhanced by bias generating the
considerable amplification of diffusion in comparison with free
diffusion. One can expect that such a relationship reflects some
intrinsic features of the interdependence of diffusion and current
driven by external force.

\vspace{-20pt}
\begin{ack}
\vspace{-15pt} The authors are grateful to Romi Mankin and Marco
Patriarca for valuable discussions, and acknowledge support by
Estonian Science Foundation through Grant No. 5662.
\end{ack}


\vspace{-15pt}

\begin{thebibliography}{00}
\vspace{-15pt}

\bibitem{constantini} G.Constantini, F.Marchesoni, Europhys. Lett. 48 (1999) 491.
\bibitem{reimann2001} P.Reimann, C.Van den Broeck, H.Linke, P.H\"anggi, J.M.Rubi,
A.P\'erez-Madrid, Phys. Rev. Lett. 87 (2001) 010602.
\bibitem{reimann2002} P.Reimann, C.Van den Broeck, H.Linke, P.H\"anggi,
J.M.Rubi,
A.P\'erez-Madrid, Phys. Rev. E 65 (2002) 031104.
\bibitem{lindner2001} B.Lindner, M.Kostur, L.Schimansky-Geier, Fluct. Noise Lett. 1
(2001) R25.
\bibitem{dan} D.Dan, A.M.Jayannavar, Phys. Rev. E 66 (2002) 041106.
\bibitem{lindner2002} B.Lindner, L.Schimansky-Geier, Phys. Rev. Lett. 89 (2002) 230602.
\bibitem{risken} H.Risken, The Fokker-Planck Equation, Springer-Verlag, Berlin, 1996.
\bibitem{svoboda} K.Svoboda, P.P.Mitra, S.M.Block, Proc. Natl. Acad. Sci. USA
91 (1994) 11782.
\bibitem{schnitzer} M.J.Schnitzer, S.M.Block, Nature 388 (1997) 386.
\bibitem{visscher} K.Visscher, M.J.Schnitzer, S.M.Block,
Nature 400 (1999) 184.

\end{thebibliography}
\end{document}